\documentclass[journal]{vgtc}                     

\graphicspath{{figs/}{figures/}{pictures/}{images/}{./}} 

\usepackage{tabu}                      
\usepackage{booktabs}                  
\usepackage{lipsum}                    
\usepackage{mwe}                       
\usepackage{amsmath}

\usepackage{booktabs, multirow, siunitx}

\usepackage[normalem]{ulem}

\usepackage[svgnames]{xcolor}
\usepackage{tcolorbox}
\usepackage[commandnameprefix=always]{changes}
\usepackage{siunitx}
\usepackage[pagebackref,bookmarks]{hyperref}

\onlineid{1490}

\vgtccategory{Research}

\vgtcpapertype{theory/model}

\title{Tell Me Without Telling Me:\\ Two-Way Prediction of Visualization Literacy and Visual Attention}

\author{%
\authororcid{Minsuk Chang}{0009-0007-5088-8991},
\authororcid{Yao Wang}{0000-0002-3633-8623},
\authororcid{Huichen Will Wang}{0009-0007-5941-4047},
Yuanhong Zhou,\\ 
\authororcid{Andreas Bulling}{0000-0001-6317-7303}, and
\authororcid{Cindy Xiong Bearfield}{0000-0002-1451-4083}
}

\authorfooter{
\item
    Minsuk Chang, Yuanhong Zhou, and Cindy Xiong Bearfield is with Georgia Tech.
    E-mail: \{minsuk, yzhou842, cxiong\}@gatech.edu
\item
    Yao Wang and Andreas Bulling is with University of Stuttgart.
    E-mail: \{yao.wang, andreas.bulling\}@vis.uni-stuttgart.de
\item
    Huichen Will Wang is with University of Washington.
    E-mail: wwill@cs.washington.edu
}

\abstract{%
Accounting for individual differences can improve the effectiveness of visualization design. 
While the role of visual attention in visualization interpretation is well recognized, existing work often overlooks how this behavior varies based on visual literacy levels.
Based on data from a 235-participant user study covering three visualization tests (mini-VLAT, CALVI, and SGL), we show that distinct attention patterns in visual data exploration can correlate with participants' literacy levels: While experts (high-scorers) generally show a strong attentional focus, novices (low-scorers) focus less and explore more. 
We then propose two computational models leveraging these insights: \textit{Lit2Sal} -- a novel visual saliency model that predicts observer attention given their visualization literacy level, and 
\textit{Sal2Lit} -- a model to predict visual literacy from human visual attention data. 
Our quantitative and qualitative evaluation demonstrates that Lit2Sal outperforms state-of-the-art saliency models with literacy-aware considerations.
Sal2Lit predicts literacy with 86\% accuracy using a single attention map, providing a time-efficient supplement to literacy assessment that only takes less than a minute.
Taken together, our unique approach to consider individual differences in salience models and visual attention in literacy assessments paves the way for new directions in personalized visual data communication to enhance understanding.
}


\usepackage{mathptmx}                  

\begin{document}

\teaser{
   \centering
    \includegraphics[
    width=\linewidth,
    alt={Predicted saliency maps for high vs low literacy on sample mini VLAT, CALVI, SGL items; experts focus, novices diffuse.}
    ]{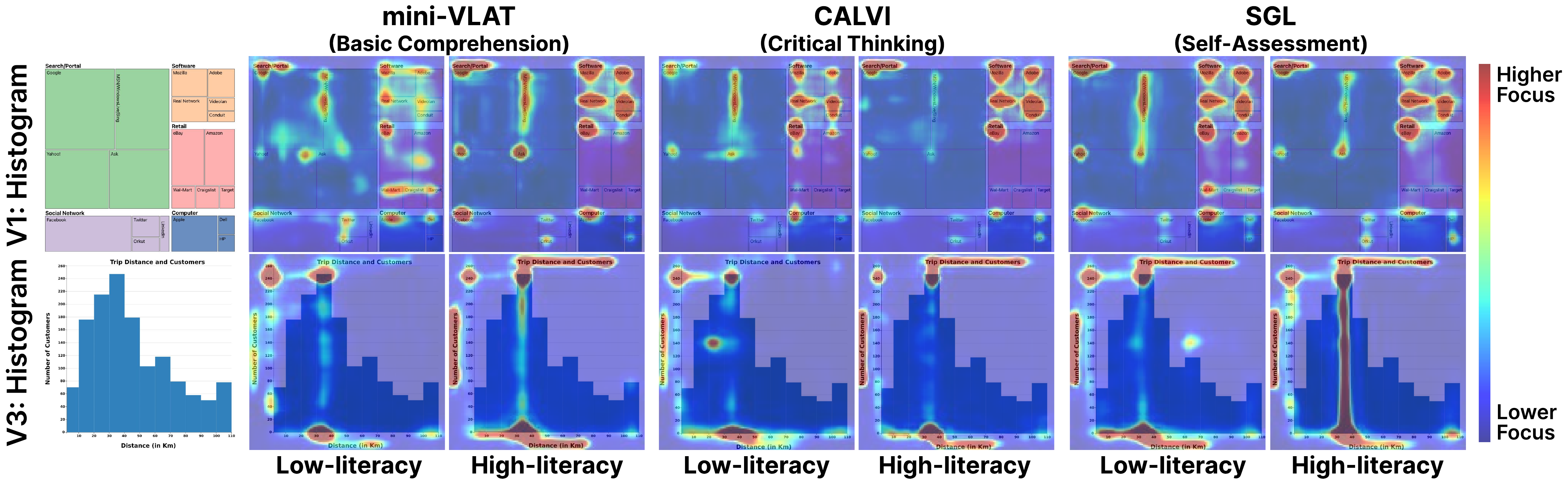}
    \caption{Predicted visual saliency map from Lit2Sal model on three literacy assessment test (mini-VLAT, CALVI, SGL) scores for treemap and histogram. High-literacy group represents the top 20\% on each test, and Low-literacy group represents the bottom 20\%. The saliency map for high scores on mini-VLAT and CALVI focuses more on the title and axis labels than low scores, also showing more focused attention. Refer to \autoref{sec:saliency_prediction_model} for further details.
    }
\label{fig:saliency_prediction}
}


\firstsection{Introduction}
\maketitle

Visualization can guide users to notice key patterns in data.
Yet what counts as the 'right' design often depends on who is looking.
People interpret data on a deeply personal level~\cite{suresh2021beyond, peck2019data}, thus visualization design effectiveness can depend on their analytic tasks~\cite{lee2022affective}, their goals~\cite{quadri2024you}, and most importantly, their visualization literacy~\cite{nobre2024reading, mini-vlat}.

Existing research has demonstrated that we can improve visualization design by understanding where people look in a visualization~\cite{graphicDesignImportance}. 
Researchers have built saliency models to predict which parts of a visualization are most likely to attract a viewer’s attention~\cite{DVSaliencyModel2017Matzen, bylinskii2016should, bylinskii2018different, gazeUncertainty, scannerDeeply}.
These models have proven valuable in supporting visualization and tool design, such as attention-aware UI~\cite{muller2022designing}, chart compression~\cite{graphicDesignImportance}, and image quality evaluations~\cite{tong2010full}.

However, existing models still largely assume universal viewing patterns among people and overlook differences driven by cognitive abilities such as visualization literacy~\cite{toker2013individual, national2000people}.
We posit that \textbf{individuals with varying levels of visualization literacy interpret visualizations through distinct viewing patterns and can be effectively captured by models of their visual attention.}
Therefore, we argue that we can improve existing saliency models by accounting for the unique patterns in viewers' visual attention depending on their literacy levels. 

In this paper, we conduct  user study (N = 235) using three established literacy tests: the mini Visualization Literacy Assessment Test (mini-VLAT) ~\cite{mini-vlat}, the Critical thinking Assessment for Literacy in VIsualizations (CALVI)~\cite{ge2023calvi}, and the Subjective Graph Literacy assessment (SGL)~\cite{garcia2016measuring}.
We recorded participants' responses and attention maps generated using the established BubbleView~\cite{bubbleView} technique, which approximates gaze tracking with mouse clicks. 
\textbf{Our analyses reveal striking attention differences between high- and low-literacy groups in each assessment}: 
Experts focus on specific regions of visualizations, creating concentrated `hot spots' that reflect their targeted attention.
In contrast, the viewing patterns of novices tended to be more distributed, with less intense focal points. 
This finding provides strong evidence that performance on visualization literacy tests does correlate with distinct visual attention patterns when interpreting visualizations.
Informed by these insights, we then introduce two novel saliency models for literacy-aware attention prediction (\textbf{Lit2Sal}) and visual attention-based visualization literacy prediction (\textbf{Sal2Lit}).

\textbf{Lit2Sal} extends the state-of-the-art saliency model VisSalFormer~\cite{salchartqa} to predict where people look in a visualization while accounting for their literacy levels, across three literacy assessment categories: mini-VLAT~\cite{mini-vlat} for visualization comprehension, CALVI~\cite{ge2023calvi} for critical thinking abilities, and SGL for self-perceived literacy, both individually and holistically.
By integrating these measures, our model generates distinct saliency predictions for novices and experts.

\textbf{Sal2Lit} can predict visualization literacy levels for all three tests with an average accuracy of 86\% with a human attention map from only one visualization.
If we expand the input data to include attention maps from three visualizations, our model accuracy increases by over 87\% (mini-VLAT) to 90\% (CALVI, SGL). 
This time-efficient proxy supplements existing literacy assessments and reveals visual processes that differentiate novices and experts in visual data exploration.

\vspace{2mm}
\noindent \textbf{Contributions:} 
We conducted a crowdsourced study and a series of analyses, discovering how different aspects of visualization literacy (e.g., basic comprehension, critical thinking, and self-assessed ability) correlate with viewers’ gaze patterns. 
Our literacy-to-saliency model, Lit2Sal, enhances existing saliency models by incorporating individual differences and aligning attention predictions with a viewer’s visualization literacy level.
Conversely, our saliency-to-literacy model, Sal2Lit,  takes a perception-driven approach to reframe how visualization literacy can be assessed through observed visual attention patterns.
Together, they advance our fundamental understanding of visual patterns in visual data exploration.
By embracing personalization, these models empower researchers and practitioners to design personalized visualizations and efficiently estimate literacy levels.

\begin{figure*}[th]
    \centering
    \includegraphics[
    width=\linewidth,
    alt={Model architecture of Lit2Sal and Sal2Lit. Left panel shows the BubbleView interface with blurred chart and click-based attention capture; middle panel illustrates the Lit2Sal model with Swin-transformer, BERT, and literacy-conditioned fusion; right panel depicts Sal2Lit’s feedforward neural network predicting literacy levels from attention features.}
    ]{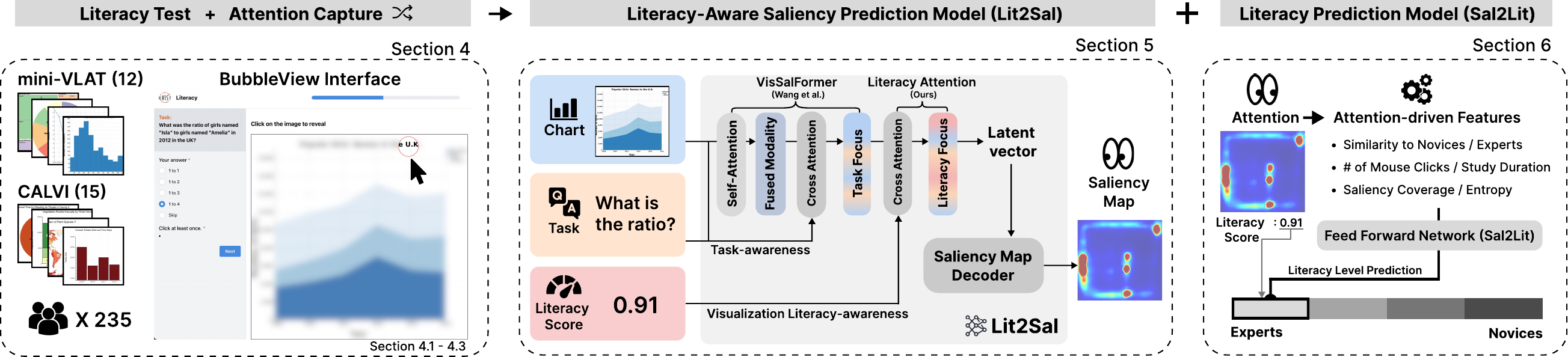}
    \caption{Overview of our study from data collection to model experiments. We first collected attention data from 235 participants using BubbleView~\cite{bubbleView} while they solved 27 visualization literacy test questions consisting of mini-VLAT~\cite{mini-vlat} and CALVI~\cite{ge2023calvi}. The data was used to test hypotheses in \autoref{sec:result_h1}-\ref{sec:result_h3} and train two models: Lit2Sal (\autoref{sec:saliency_prediction_model}), and Sal2Lit (\autoref{sec:literacy_level_prediction_model}).}
    \label{fig:flow}
    \vspace{-3mm}
\end{figure*}
\section{Related Work}

\subsection{Visual Attention in Visualizations}
Where do people look in a visualization?
Cognitive and vision scientists have long studied human attention to understand how visual features such as color, contrast, and orientation can guide attention patterns~\cite{treisman1980feature, itti1998model}. 
With the rise of large-scale eye-tracking datasets and advances in machine learning, researchers have developed more sophisticated models to predict human attention patterns across images, text, and data visualizations. 
This research identifies salient regions that most likely attract human attention~\cite{shanmuga2015eye, bonhage2015combined, conklin2016using}.
Saliency models serve as critical tools for evaluating the visual hierarchy in attention allocation in scene interpretation~\cite{itti1998model} and object identification~\cite{borji2015salient}. 

While conventional saliency models can efficiently predict human attention patterns in natural scenes, they fail to account for the specialized perceptual and cognitive principles that govern our processing of data visualizations~\cite{franconeri2021science, correll2012comparing, polatsek2018exploring, knittel2024gridlines}.
This results in inaccurate predictions of attention patterns with visualized data. 
To address this, visualization researchers have developed saliency prediction models tailored to visualizations~\cite{DVSaliencyModel2017Matzen, bylinskii2016should}. 
However, these early attempts primarily relied on manually crafted rules based on pixel values, which can lose generalizability when the visualization gets complex and diverse.
Recent advancements leverage deep learning techniques and extensive eye-tracking datasets in both natural images~\cite{lou2022transalnet} and visualizations~\cite{fosco2020predicting, scannerDeeply} to improve prediction accuracy.
Furthermore, innovative eye-tracking alternatives such as webcam-based tools~\cite{webgaze} and mouse interactions~\cite{importAnnot, bubbleView, codechart} have enabled researchers to collect large-scale saliency data, overcoming the practical limitations associated with conventional eye-tracking approaches.
Advances in visual attention modeling also considered factors such as task relevance~\cite{salchartqa} and gaze path sequences~\cite{scanpath}, leading to more nuanced models. 
These approaches typically aggregate saliency data from all participants to define a general consensus-based ground truth. 
We supplement these approaches to further improve visualization saliency models by considering individual data literacy levels in visual data interpretation, 

Alongside the model development, various datasets have been collected to explore and analyze viewer saliency patterns in visualizations more systematically~\cite{burch2017eye, kurzhals2014evaluating, scannerDeeply}. 
Borkin et al.~\cite{borkin2015beyond} collected an eye-tracking dataset with 393 visualizations from the MASSVIS~\cite{borkin2013makes} dataset. 
Shin et al.~\cite{scannerDeeply} crowdsourced saliency maps over 10,000 visualizations via webcam-based eye tracking. 
However, both datasets were collected in a task-agnostic setting, where saliency data was collected without considering the top-down influence on visual attention.
To supplement the lack of task-specific saliency data, Gomez et al.~\cite{gomez2016fauxvea} created a dataset of 20 visualizations with task-driven saliency, covering analytic tasks, including value retrieval in bar charts and outlier point detection in scatterplots. 
Polatsek et al.~\cite{polatsek2018exploring} analyzed how low-level tasks affect saliency through an eye-tracking study with 30 visualizations.
Most recently, with a mouse-based proxy BubbleView~\cite{bubbleView}, Wang et al.~\cite{salchartqa} collected the first large-scale saliency dataset containing task-relevance regions using 3,000 visualizations from ChartQA~\cite{chartQA} dataset. 
However, no existing dataset has linked saliency patterns with visualization literacy, despite evidence that viewing patterns can predict cognitive abilities~\cite{conati2020comparing}.
We address this gap by collecting saliency maps along with three literacy assessments, annotating literacy test performance from VLAT~\cite{mini-vlat}, CALVI~\cite{ge2023calvi}, and SGL~\cite{garcia2016measuring} to provide insights into the effect of individual differences in saliency modeling.

\subsection{Visualization Literacy Assessments} \label{sec:relwork2}
\label{sec:visualization_literacy}
Visualization literacy has long been a critical area of interest in the visualization community~\cite{boy2014principled}.
Early assessments of visualization literacy often relied on unstructured methods, primarily designed for specific domains, which involved manually curated visualization-question pairs~\cite{maltese2015data, borner2016investigating, kwon2016comparative, ruchikachorn2015learning, wang2020cheat}. 
For instance, Schönborn et al.~\cite{Schönborn2006} analyzed biochemistry students’ ability to understand external representations and offer insights into domain-specific literacy. 
Baker et al.~\cite{baker2001toward} and Huron et al.~\cite{Huron2014Constructing} measured the visualization literacy of participants by making them create or update the charts. 
Rodrigues et al.~\cite{rodrigues2021questions} explored how novices attempt to interpret visualizations by letting them ask questions rather than responding to predefined ones. 

However, these domain-specific and exploratory methodologies limited the generalizability of research findings and comparability across different studies.
To address this gap, Lee et al.~\cite{lee2016vlat} proposed VLAT, a standardized test assessing basic chart comprehension across diverse visualization types and tasks.
Garcia et al.~\cite{garcia2016measuring} created a standardized self-assessment inventory of visualization literacy based on chart familiarity. 
In addition to evaluating chart comprehension and familiarity, researchers also recognized that literacy could be assessed through higher levels of thinking, such as critical reflection of visualization takeaways and evaluations of argument quality from visualizations~\cite{burns2020evaluate}. 
For example, to evaluate the critical thinking process of visualization literacy, Ge et al.~\cite{ge2023calvi} created CALVI by adding trick questions where misinformation is deliberately included. 
Quadri et al.~\cite{quadri2024you} assessed the viewer's high-level chart comprehension by qualitatively comparing the takeaways with theme analysis.
Adar and Lee~\cite{adar2020communicative} introduced a learning objectives framework to evaluate how viewers comprehend communicative visualizations matching the designer's goal.
These assessments were frequently adopted in other research to assess general visualization literacy~\cite{lee2019correlation, donohoe2020data}, understand barriers better~\cite{nobre2024reading, choe2024enhancing}, or benchmark large language models' visualization literacy~\cite {bendeck2024empirical, pandey2025benchmarking, hong2025llms}. 

Researchers also considered enhancing the reliability and efficiency of these standardized tests by slightly tuning their structures.  
For example, mini-VLAT, a subset of VLAT~\cite{lee2016vlat}, was developed with comparable diagnostic capability. 
Cui et al.~\cite{cui2023adaptive} also devised a compressed version of CALVI, which reduced the average test questionnaires by half~\cite{cui2023adaptive}. 
Despite these efforts, assessing literacy remains time-intensive, particularly across multiple cognitive levels in Bloom’s Taxonomy~\cite{forehand2010bloom}, from basic comprehension to critical thinking.
This further motivated us to estimate literacy levels efficiently. 

Although standardized tests provide individual score-based assessments, researchers typically group people to understand their common aspects better. For example, in the field of education, Persky et al.\cite{expertiseEducation2} proposed a framework comprising five stages of expertise, ranging from beginner to expert, drawing inspiration from the Dreyfus model~\cite{dreyfus2004five}, which similarly categorizes adults’ skill acquisition into five distinct phases. Reding et al.~\cite{expertiseEducation1} aimed to split the participants into four quantiles but employed a binary classification of experts and novices due to the limited dataset size. For visualization literacy, proficiency has been self-assessed using three categories: novice, intermediate, and expert ~\cite{van2024understanding}. Building on this literature, we design a predictive model to assess literacy among two to five levels. Drawing from various definitions, we also use the terms experts and novices to refer to individuals with the highest and lowest visualization literacy, respectively, as measured by test scores.

\section{Hypotheses}
To understand how visualization literacy relates to visual attention, we formulate three hypotheses that examine this relationship in progressive stages. First, we discover how different aspects of literacy are measured and interrelated.
Existing assessments of visualization literacy vary in the types of visualizations and questions they employ~\cite{mini-vlat, lee2016vlat, ge2023calvi, garcia2016measuring}, each capturing different aspects of literacy. 
We focus on three widely used assessments with different types of assessments: mini-VLAT~\cite{mini-vlat} (basic comprehension), CALVI~\cite{ge2023calvi} (critical thinking), and SGL~\cite{garcia2016measuring} (self-assessment), and hypothesize that 
\begin{tcolorbox}[boxsep=2pt, left=2pt, right=2pt, top=2pt, bottom=2pt]
\textbf{H1:} The distribution of participants’ scores will differ across the assessments, reflecting the distinct aspects of literacy that each test is designed to measure.
\end{tcolorbox}
Then, we analyze how attention pattern differs between groups of participants with different literacy levels.
In learning science, experts and novices are known to exhibit different cognitive processes, particularly in how they retrieve and organize relevant information~\cite{national2000people}.
Given that visualization viewing patterns are task-dependent\cite{salchartqa, chang2025grid} and can vary significantly based on individual characteristics such as verbal working memory\cite{toker2013individual}, we hypothesize that 
\begin{tcolorbox}[boxsep=2pt, left=2pt, right=2pt, top=2pt, bottom=2pt]
\textbf{H2:} Visualization experts and novices will exhibit distinct viewing patterns, leading to differences in their generated attention maps.
\end{tcolorbox}

Considering both \textbf{H1} and \textbf{H2} with an assumption that we could build a bidirectional mapping between visualization literacy level and visual saliency, we finally hypothesize
\begin{tcolorbox}[boxsep=2pt, left=2pt, right=2pt, top=2pt, bottom=2pt]
\textbf{H3:} Visualization literacy and viewing patterns will be correlated, such that high-literacy participants exhibit viewing behaviors similar to experts, while low-literacy participants resemble novices.
\end{tcolorbox}

Finding evidence for these hypotheses would give us a clear view of how visualization literacy and viewing patterns correlate, providing a foundation for designing more adaptive and user-aware visualizations.
\section{Experiment}
We conducted a crowdsourcing study to evaluate our hypotheses.
\autoref{fig:flow} shows study results being used to build our literacy-to-saliency prediction model (Lit2Sal) and saliency-to-literacy prediction model (Sal2Lit). The study was deployed through Revisit~\cite{revisit} and Prolific.

\subsection{Participants}
Using the 
pilot study data with 40 participants, we conducted a power analysis to identify the smallest effect size across all features listed in \autoref{sec:result_h3}, which had a Cohen's $f$ of 0.057.
Following the power analysis with G*Power, we recruited 235 participants to detect an effect with 95\% power at $\alpha$ = 0.05. 
The participants had a mean age of  $M_{age}$ = 38.73  years ($\sigma_{age}$ = 13.03), with 89 identifying as female. 
Regarding education, the plurality (89) held a bachelor’s degree or higher, followed by 67 with a high school diploma or equivalent and 66 with a master’s degree. 
Regarding visualization experience, 83 participants reported occasionally reading data visualizations, while 59 did so frequently. Additionally, 43 occasionally created visualizations for work or hobbies, and 30 reported often creating data visualizations.
All participants were compensated at a rate of \$12 per hour.

\begin{figure*}[t]
\centering\includegraphics[
    width=\linewidth,
    alt={Multiple Correspondence Analysis results with three components shown as an elbow point in the eigenvalue chart. Questions are color-coded by the component they contribute most to.}
    ]{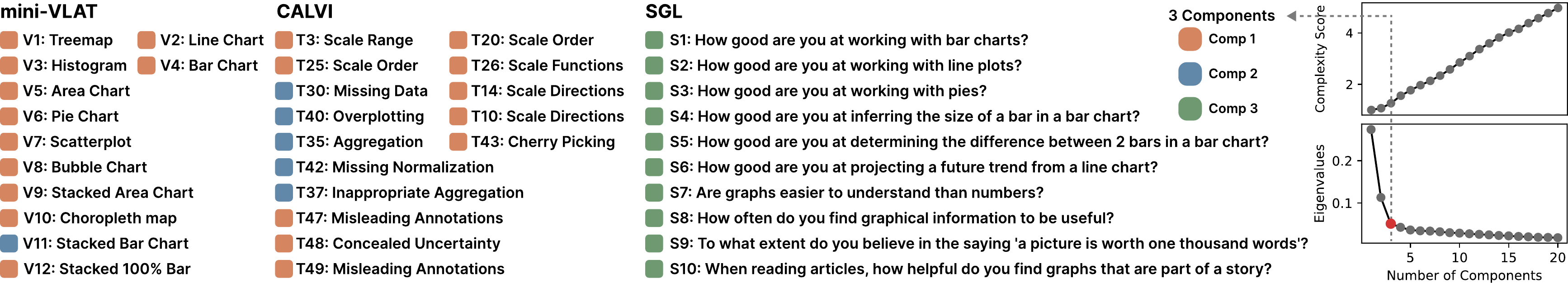}
    \caption{Multiple Correspondence Analysis (MCA) results indicate three distinct components forming an elbow point in the component-eigenvalue chart. Each question is color-coded based on the component it contributes to the most.}
    \label{fig:factor}
    \vspace{-4mm}
\end{figure*}

\subsection{Methodology}
\noindent \textbf{Visualization Literacy Tests:}
Since visualization literacy can be assessed in multiple ways (\autoref{sec:visualization_literacy}), we employed three complementary approaches for a more comprehensive measurement: mini-VLAT~\cite{lee2016vlat} (basic literacy), abridged CALVI~\cite{ge2023calvi} (critical thinking), and SGL~\cite{garcia2016measuring} (self-assessment). 

We used the mini-VLAT~\cite{mini-vlat} to assess basic comprehension to ensure the study runs for a manageable length.
For critical thinking, because the full-length CALVI~\cite{ge2023calvi} consists of 47 trick questions and requires over 45 minutes to complete, we selected a 15-item subset with the highest discriminability, following the authors’ recommendation in subsequent work by Cui et al.~\cite{cui2023adaptive}.
This combination of mini-VLAT and abridged CALVI contains both questions with and without misinformation, which Ge et al.~\cite{ge2023calvi} have demonstrated to support participants’ critical thinking process. 
SGL, being a standalone, short self-assessment, was administered in its entirety. 

\vspace{2mm}
\noindent \textbf{Capturing Viewing Patterns through Attention Proxy}
We also showed each visualization on mini-VLAT and abridged CALVI with the BubbleView~\cite{bubbleView} interface to collect the viewing pattern when taking the tests. 
While traditional eye-tracking hardware is effective, it can be cumbersome for collecting large-scale attention data. Recent research has validated mouse-based proxies as reliable alternatives for capturing attention from images~\cite{turkeyes, bubbleView, importAnnot, codechart, webgaze, zoommaps, chang2025grid}.
Therefore, we used a mouse-based proxy, BubbleView~\cite{bubbleView}, for data collection, especially suited for capturing intentional attention. The bubble size was initially set to 32px and then adjusted based on each participant’s screen. This ensured consistent bubble sizes across different images and display resolutions, aligning with the size of human foveal vision~\cite{salchartqa, bubbleView}. From the collected stream of click records, we generated a continuous 2D attention map by applying a Gaussian blur to the bubbles, using a kernel with a sigma of 19px, approximately the same as 1 degree of visual angle~\cite{bylinskii2018different, bubbleView}.
Our pilot study (40 participants) showed that 95\% of participants answered the literacy questions within 85.02 seconds per chart.
Therefore, we displayed each visualization to a participant for 90 seconds, allowing them ample time to visually explore and answer the mini-VLAT and CALVI literacy questions.
This time limit also served as an attention check to encourage participants to efficiently engage with each question.

\subsection{Procedure}
The study was conducted as a between-subject experiment. 
After consenting to the study, participants self-assessed their visualization literacy with the SGL~\cite{garcia2016measuring} test consisting of 10 questions with a 1-6 Likert scale. 
Then, we randomly presented them with 27 mixed multiple-choice questions from mini-VLAT and CALVI.
The visualizations associated with each question were presented with a BubbleView interface, as shown in \autoref{fig:flow} (left block). 
Each participant in main study had 90 seconds to freely explore the given visualization in BubbleView and answer the associated literacy question; pilot study had no time constraints. Also there were no specific instructions given on how to explore.
After 90 seconds, they could only skip the question, but all interactions were logged.
Participants could not return to previous questions. Upon completion, they provided demographic information: age, gender, education level, and experience in data visualization.

\subsection{H1: Discriminability of Literacy Assessment Tests} \label{sec:result_h1}
We first analyzed how the three literacy assessment tests differ regarding overall score distribution and participant discriminability.

\subsubsection{Overall Score Distribution} \label{sec:score_distribution}
\begin{figure}[t]
    \centering
    \includegraphics[
    width=\linewidth, 
    alt={Score distributions of mini-VLAT, CALVI, and SGL tests, normalized to [0,1]. Left plots show histograms with skewness differences, while the right scatterplots compare mini-VLAT scores to CALVI and SGL, illustrating the Dunning-Kruger effect.}
    ]{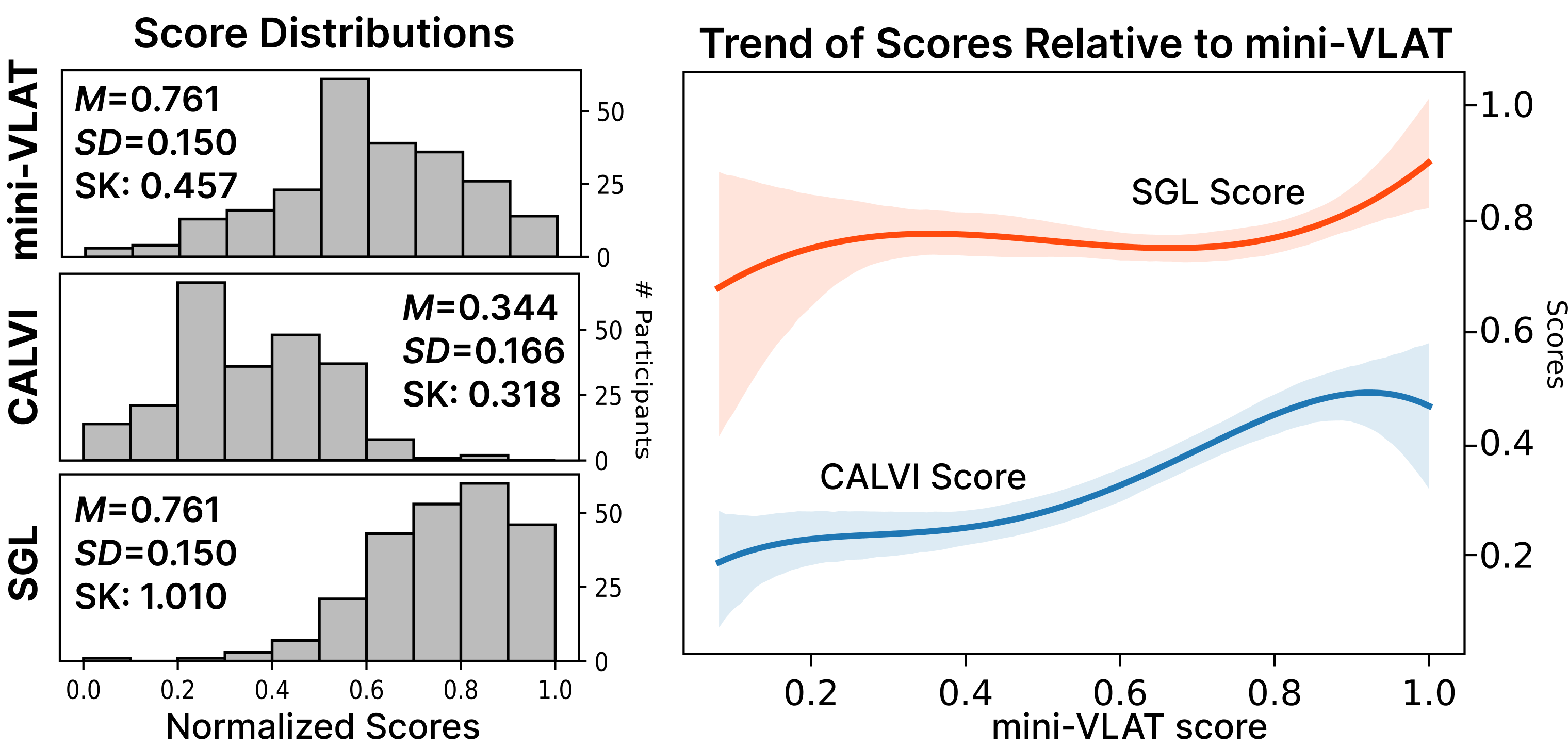}
    \caption{(left) Score distribution for each test with statistics (average, standard deviation, and Pearson's 3rd skewness). All scores are normalized within [0, 1] based on the possible lowest and highest scores for each test. (right) Relationship between mini-VLAT score and other two scores (CALVI, SGL). The order of polynomial regression lines was calculated based on the elbow-point on the $R^2$ metric.}
    \label{fig:dunning}
    \vspace{-5mm}
\end{figure}

The score distributions with basic statistics are shown in \autoref{fig:dunning}, where each score was normalized to the [0, 1] range based on its possible minimum and maximum values. For mini-VLAT, we first applied the score correction formula $R - W/(C - 1)$ (where R: correct, W: wrong, C: number of choices) to account for random guessing, as recommended by the authors~\cite{mini-vlat}, while CALVI and SGL used raw scores. Throughout the paper, we refer to the normalized scores for all three tests instead of the raw scores unless stated otherwise.

Each test score distribution resembled a normal distribution, with the mini-VLAT exhibiting the widest and most balanced spread.
Regarding the balancedness of each test, we measured Pearson's third-moment coefficient of skewness. mini-VLAT and CALVI are slightly skewed to the left and right, respectively. SGL was highly left-skewed, showing a ceiling effect.

\subsubsection{Relationships between Literacy Assessment Scores} \label{sec:discriminability}
The ceiling effect observed in SGL~\cite{garcia2016measuring} correlates with performance on the mini-VLAT~\cite{mini-vlat} with a resemblance to the Dunning-Kruger effect~\cite{kruger1999unskilled, chen2024unmasking}.
\autoref{fig:dunning} presents a regression plot of mini-VLAT scores against CALVI and SGL scores, highlighting this phenomenon.
Participants with very low or moderately high mini-VLAT performance estimated their literacy to be low in the self-assessment (SGL). In contrast, those with slightly low or exceptionally high VLAT performance estimated their literacy to be high in SGL.

\begin{figure*} [t]
\centering
    \includegraphics[
    width=\linewidth,
    alt={Attention maps for high vs low literacy on sample items; expert hotspots contrast with novice spread.}
    ]{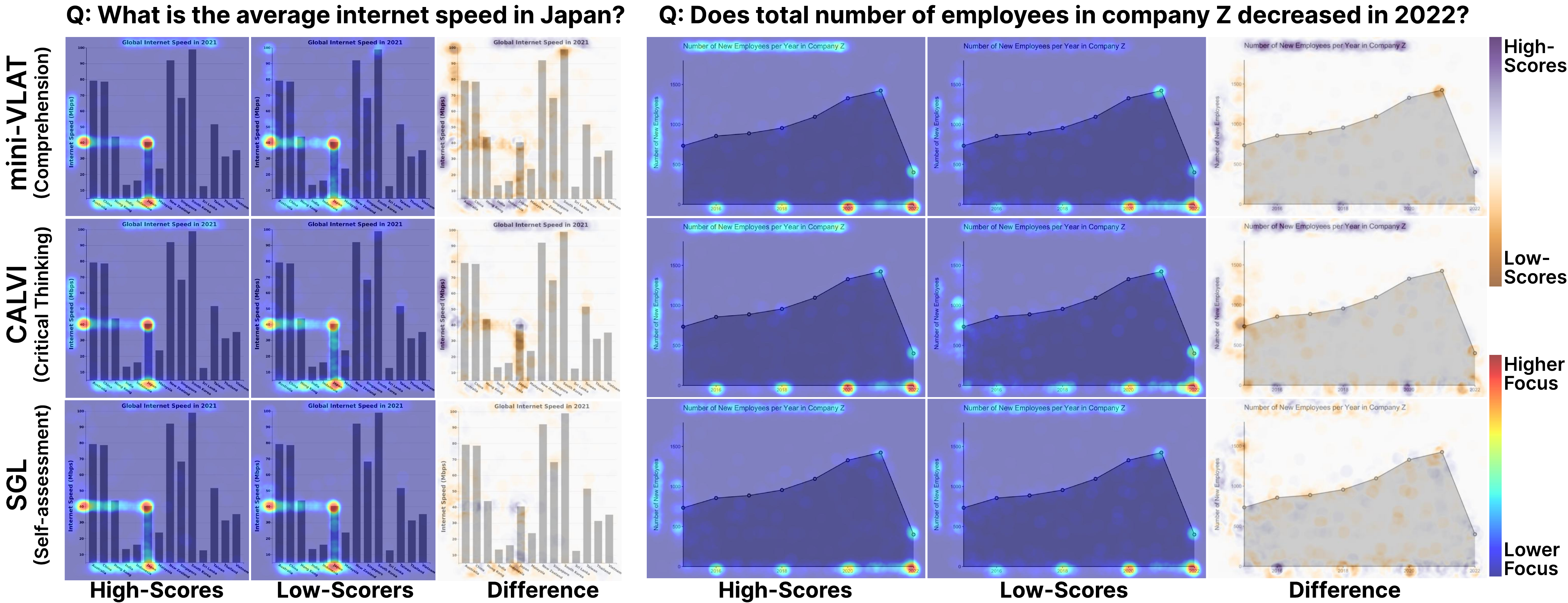}
    \caption{
    Differences in visual attention between high- and low-scorers on three literacy tests (mini-VLAT, CALVI, and SGL) with V4 (Bar Chart) and T37 (Inappropriate Aggregation) charts. High and low scorers are defined as the top and bottom 25\% of participants. In the attention maps, red regions indicate areas of higher visual attention, while blue regions indicate lower attention. The difference map highlights areas where high-performers exhibited greater attention in purple and where low-performers did in orange.
    }
    \label{fig:teaser}
    \vspace{-5mm}
\end{figure*}

We formally characterize the relationship between mini-VLAT and SGL by analyzing the variance explained in a polynomial regression model. 
As the polynomial degree increases, the model captures more variance and becomes more complex. 
To balance model fit and complexity, we identify the elbow point where $R^2$ shows minimal improvement with additional degrees. 
This occurs at degree three, indicating that a third-degree polynomial provides the optimal trade-off.

This allows us to express the relationship between SGL and VLAT performance as follows:
\begin{equation}
    \text{S} = 0.598 + 1.216 \cdot \text{V} - 2.619 \cdot \text{V}^2 + 1.708 \cdot \text{V}^3
\end{equation} 
where S refers to SGL and V to the mini-VLAT score. The extreme points where the Dunning-Kruger effect occurs are V = 0.36 and 0.67.

Similarly, we fit a polynomial regression model to describe the CALVI score with mini-VLAT. $R^2$ score had the elbow point on the fourth-degree polynomial, showing optimal trade-offs. The expressed relationship is:
\begin{equation}
    \text{C} = 0.112 + 1.267 \cdot \text{V} - 4.852 \cdot \text{V}^2 + 7.989 \cdot \text{V}^3 - 4.046 \cdot \text{V}^4
\end{equation} 
where C is the CALVI score. 
Analysis of the second derivative of the curve reveals a non-decreasing trend in C when the V ranges from 0 to 0.95.
This suggests that, compared to SGL, CALVI shows a stronger positive correlation with mini-VLAT scores, indicating a relatively similar aspect of literacy measurement.

\begin{figure}[t]
    \centering
    \includegraphics[
    width=\linewidth,
    alt={Attention maps split by correct vs. incorrect responses across mini-VLAT and CALVI. Correct responses focus more on titles, axis labels, and legends, whereas incorrect responses have scattered, less focused attention.}
    ]{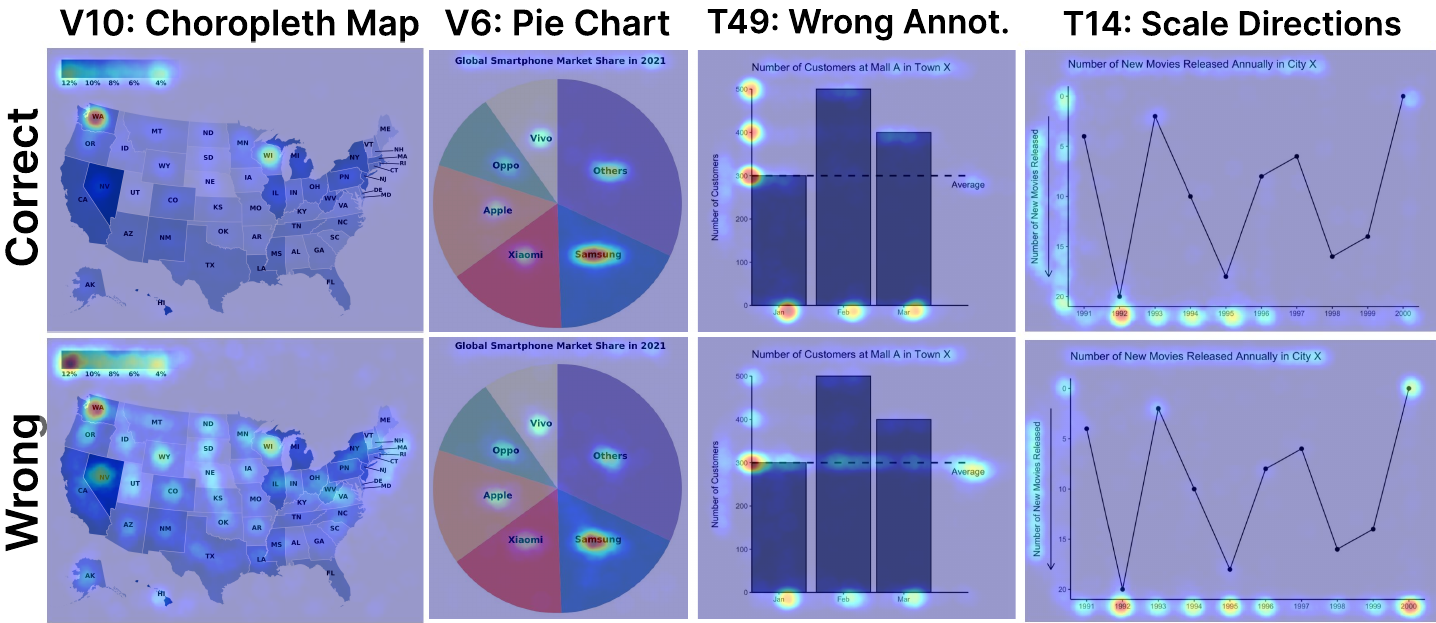}
    \caption{Averaged attention map of participants who answered the task question correctly or not. The V10 and V6 were sampled from mini-VLAT, and T49 and T14 from CALVI. The chart codes are described in~\autoref{fig:factor}.
    }
    \label{fig:correctness}
    \vspace{-5mm}
\end{figure}

\subsubsection{Multiple Correspondence Analysis (MCA)}
We conducted a multiple correspondence analysis to examine further whether the three literacy tests capture distinct aspects of literacy, as they claim.
For instance, mini-VLAT is focused on basic chart comprehension~\cite{mini-vlat}, CALVI is for critical thinking ability~\cite{ge2023calvi}, and SGL is for self-assessment~\cite{garcia2016measuring}. 
To uncover latent dimensions across the three tests, we applied MCA to item-level responses, using binary correctness for mini-VLAT and CALVI and a binarized form of SGL’s six-level Likert scale. For SGL, we binarized the response by grouping the lower three levels as `low' and the upper three as `high'.
As shown in \autoref{fig:factor}, the eigenvalues from the MCA indicate the variance explained by each additional component.
At N = 3 components, we observe an 'elbow point' where each additional component after that no longer significantly explains more variances in the model despite increasing complexity.
This motivates us to interpret this data with a three-factor model derived from MCA~\cite{revelle2015package}. 

A closer examination of the component loadings from MCA reveals that all SGL items are positioned exclusively along Component 3, with no overlap from mini-VLAT or CALVI items. 
This indicates that the self-perceived literacy measured by SGL represents a distinct construct, separate from the dimensions assessed by the other two tests. 
In contrast, mini-VLAT and CALVI items are distributed across Components 1 and 2, indicating the presence of two additional latent structures of visualization literacy. Each test appears to contribute to both components, implying that they capture overlapping but not identical aspects of comprehension and critical reasoning.
Based on the assigned items, we interpret Component 1 as standard questions and Component 2 as difficult questions. Component 2 corresponds to a set of questions discriminating the expert level, as it includes five items from CALVI reported to be the most challenging among the 15 questions~\cite{ge2023calvi}. Among mini-VLAT questions, Component 2 includes the stacked bar chart with the lowest accuracy.

\vspace{2mm}
\noindent \textbf{Takeaway:}
These analyses show that while the three literacy assessments are correlated, they each capture distinct aspects of visualization literacy, \textbf{supporting H1}.

\subsection{H2: Viewing Patterns for Experts versus Novices} \label{sec:result_h2}

We adopt a two-step analysis to compare attention maps generated based on viewing patterns between visualization experts and novices.
First, at the individual question level, we compare attention maps associated with \textbf{correct} versus \textbf{incorrect} responses to determine whether viewing patterns differ based on response accuracy. Second, we aggregate attention maps across multiple questions for each participant to examine whether overall high-scoring individuals (i.e., \textbf{experts}) exhibit systematic differences in viewing patterns compared to low-scoring individuals (i.e., \textbf{novices}) across all three literacy assessments. The quantitative results are summarized in \autoref{tab:h2_quant}.

\begin{table}[t]
    \centering
    \resizebox{\linewidth}{!}{
   \begin{tabular}{ll|cc|ccc}
    \toprule
    \textbf{Category} & \textbf{Group} & $\textbf{Coverage}$ & $\textbf{IoU}_{\textbf{Pair}}$ & $\textbf{IoU}_{\textbf{Title}}$ & $\textbf{IoU}_{\textbf{Labels}}$ & $\textbf{IoU}_{\textbf{Legend}}$ \\
    \midrule
    \multirow{2}{*}{Accuracy} 
        & Correct     & \textbf{0.045} & \textbf{0.320} & \textbf{0.074} & \textbf{0.098} & \textbf{0.126} \\
        & Incorrect   & 0.040          & 0.245         & 0.053         & 0.084         & 0.096 \\
    \midrule\midrule
    \multirow{2}{*}{mini-VLAT}  
        & Experts     & \textbf{0.045} & \textbf{0.299} & \textbf{0.090} & \textbf{0.128} & \textbf{0.124} \\
        & Novices     & 0.039          & 0.239         & 0.042         & 0.060         & 0.094 \\
    \midrule
    \multirow{2}{*}{CALVI} 
        & Experts     & \textbf{0.046} & \textbf{0.294} & \textbf{0.092} & \textbf{0.119} & \textbf{0.123} \\
        & Novices     & 0.038          & 0.243         & 0.040         & 0.057         & 0.094 \\
    \midrule
    \multirow{2}{*}{SGL} 
        & Experts     & 0.041          & \textbf{0.260} & 0.055         & 0.087         & 0.102 \\
        & Novices     & \textbf{0.045} & 0.250         & \textbf{0.068} & 0.086         & 0.113 \\
    \bottomrule
\end{tabular}
    }
    \caption{Average metrics by accuracy and literacy groups. Coverage reflects the wideness of attention maps. IoU is computed for within-group pairs (IoU$_\text{Pair}$) or with specific chart elements (title, labels, legend). Bold indicates significance (p<0.001, pairwise $t$-tests). SGL’s IoU with labels and legend was not significant (p>0.05).}
    \label{tab:h2_quant}
    \vspace{-5mm}
\end{table}

\subsubsection{Difference of Attention by Question Accuracy} \label{sec:task_correctness}
We first qualitatively analyzed how the visual attention maps differ based on the individual correctness of the visualization literacy test questions. 
\autoref{fig:correctness} shows that the \textbf{attention maps associated with correct answers were generated by participants who were more selectively focused}, attending to fewer regions but with high intensity. 
Additionally, participants who answered CALVI questions incorrectly were more likely to attend to misinformation factors, such as incorrect annotations or peaks.
We then measured the spread of attention with Saliency Coverage~\cite{salchartqa} and within-group alignment with pairwise IoU~\cite{boggust2022shared}. 
Saliency Coverage quantifies how widely a participant’s attention is distributed (e.g., percentage of nonzero pixels in their attention map), while IoU measures the spatial overlap between two maps (e.g., intersection over union of attended regions).
\textbf{Correct group exhibited a slightly higher coverage and much higher pairwise IoU}, indicating greater consistency and overlap in their attention patterns.
This pattern makes sense because BubbleView~\cite{bubbleView} reveals chart details through a blurred image, strategically identifying task-relevant areas while minimizing exploration of irrelevant regions offers a clear advantage under time constraints.
In contrast, attention maps linked to incorrect answers tend to show low overlap between answers even though having a small coverage, indicating a higher variance of attention. 

We also observed differences in the \textbf{types of regions} associated with high attention between maps generated by correct versus incorrect responses. 
After manually splitting the charts into segments containing a title, axis label, and legend, we calculate the IoU with attention maps from correct/incorrect groups. 
Attention maps from the correct answers showed a greater focus on all three elements than those from the incorrect group, suggesting that accurate answers relied more on contextual information.
We interpret these results as participants needing to remain aware that the current plot may contain misinformation from the mixed task orders from mini-VLAT and CALVI. 
Such awareness is also reflected in the y-axis of the bar and line charts in \autoref{fig:correctness}.

\begin{figure}[t]
    \centering
    \includegraphics[
    width=\linewidth,
    alt={Feature importance comparison between linear regression coefficients and Sal2Lit model using Integrated Gradients. Darker colors indicate higher importance, with statistical significance denoted by *, **, ***.}
    ]{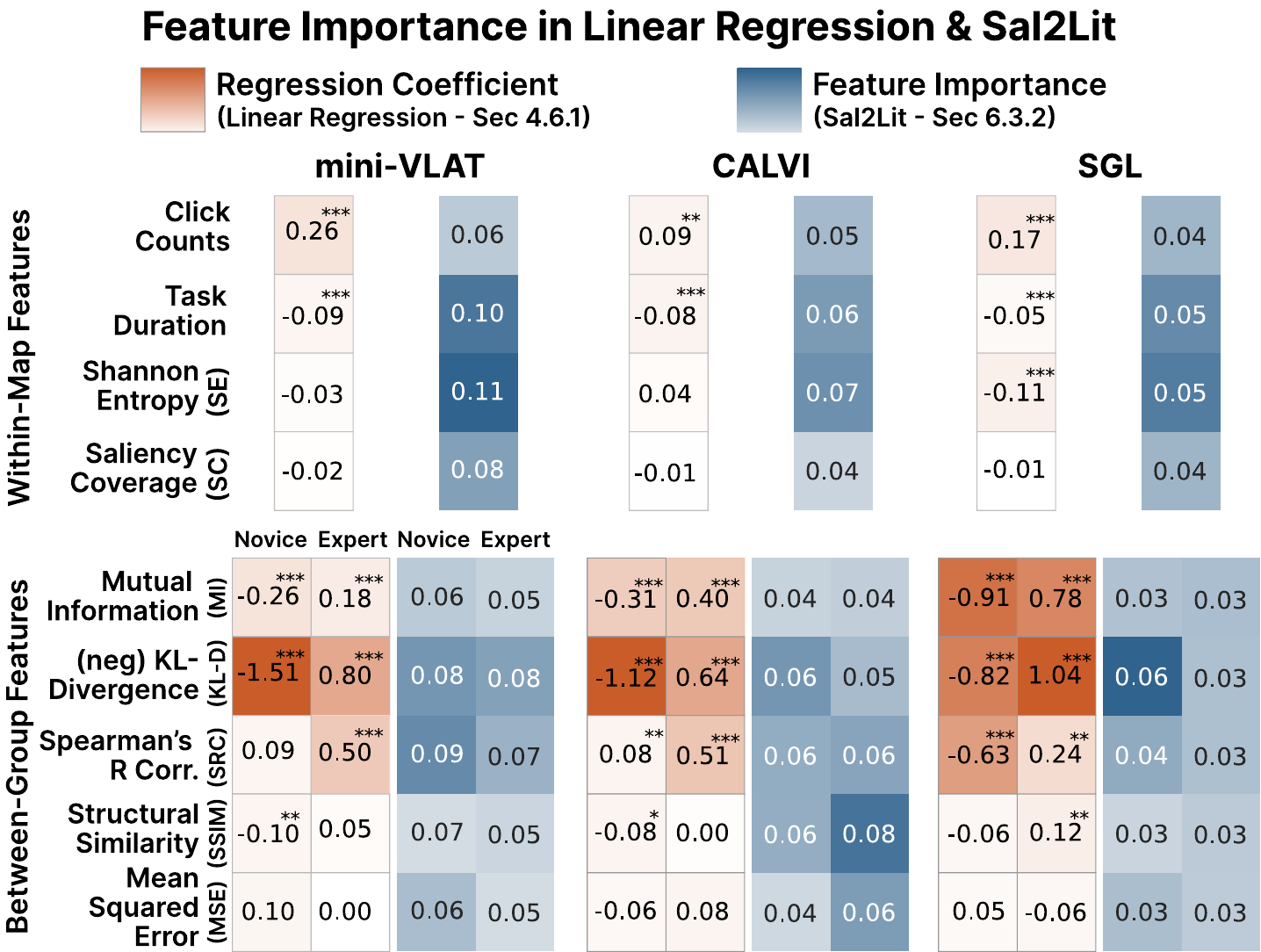}
    \caption{
    Relative feature importance from linear regression (\autoref{sec:linear_regression}) and Sal2Lit model (\autoref{sec:sal2lit_importance}). Darker color represents higher importance in both colors. The linear regression's red color represents the coefficients' absolute value with a corresponding significance level (*: p<0.05, **: p<0.01, ***: p<0.001), and blue represents the Sal2Lit model's feature importance calculated with Integrated Gradients~\cite{mudrakarta2018did}. 
    }
    \label{fig:regression}
    \vspace{-5mm}
\end{figure}

\subsubsection{Difference of Attention by Individual Literacy Scores} \label{sec:regression}
Moving beyond one-off question accuracy, we analyze participants’ overall literacy scores to identify more generalizable attention trends across literacy levels.
We sampled the top 25\% and bottom 25\% participants as experts and novices on three literacy tests, following Van et al.~\cite{expertiseEducation1}'s quartile-based grouping of expertise.

As shown in \autoref{fig:teaser}, there was a notable difference in primary focus areas across literacy levels.
Similar to our question-level attention analysis comparing correct and incorrect responses, the expert group of all literacy tests had higher pairwise IoU than the novice group, showing highly overlapping focus.
On the other hand, novice groups in all three tests had low pairwise IoU, showing a
viewing pattern of spatial exploration. 
Also, the expert group of mini-VLAT and CALVI scores focused more on contextual elements (title, labels, legends), represented with higher IoU. 
In mini-VLAT and CALVI, novices also had low IoU against textual elements, showing more attention to graphical elements where data is embedded.
However, SGL showed a different attention pattern, where novice participants focused more on the title than experts. 
Also, no significant difference in attention was found on labels and legends.
This aligns with the findings in \autoref{sec:result_h1} that SGL measures a distinct dimension of literacy compared to mini-VLAT and CALVI.

\vspace{2mm}
\noindent \textbf{Takeaway:}
Taken together, these findings \textbf{support H2}, suggesting that visual attention differs between novices and experts, driven by the distinct viewing patterns observed between correct and incorrect groups.

\subsection{H3: Correlating Literacy and Viewing Patterns} \label{sec:result_h3}
To explore the relationship between visualization literacy and attention maps generated from viewing patterns, we described participant behaviors and quantified the attention maps via two complementary categories of predictive variables:

\vspace{2mm}
\noindent \textbf{Within-Map Descriptive Features:}
We first adopt unitary features that describe the viewers' attention during the task.
These are metrics extracted from individual attention maps, such as click count, task duration, Shannon entropy (SE), and saliency coverage (SC). 
They capture a participant’s attention distribution and focus within a single visualization, following methods from Wang et al.~\cite{salchartqa}. Higher SE and SC values indicate a more broad and dispersed attention map, reflecting less focused viewing patterns. Lower values indicate more concentrated attention.

\vspace{2mm}
\noindent \textbf{Between-Group Comparative Features:}
Within-map features may only partially explain visual attention as they do not represent how the viewer's attention relates to certain groups' common behavior.
Therefore, we also aim to evaluate how each participant’s attention map aligns with others across different groups. We group the comparison maps by \textit{literacy level} (expert vs. novice) and \textit{task correctness} (correct vs. wrong). We apply five similarity metrics for each group to compare continuous attention maps.

They include: Mutual Information (MI)~\cite{viola1997alignment}, Kullback-Leibler Divergence (KL-D)~\cite{kldivergence}, Spearman's Rank Correlation (SRC)~\cite{spearman}, Structural Similarity (SSIM)~\cite{ssim}, and Mean Squared Error (MSE)~\cite{tong2010full} derived from Chang et al.~\cite{chang2025grid}. These metrics capture different similarity aspects: MI and KL-D quantify information-theoretic divergence, SRC captures monotonic spatial correlation, SSIM assesses structural layout similarity, and MSE measures raw pixel-wise difference. For KL-D, we kept the group's averaged attention as the true probability distribution.
By computing these similarity scores between each individual's maps and the aggregated version from high- and low-literacy groups, we can assess whether certain patterns are more diagnostic of literacy levels. 

Together, these two categories offer a comprehensive and explainable mapping between attention patterns and literacy to test our hypothesis.

\begin{figure}[t]
    \centering
    \includegraphics[
    width=\linewidth,
    alt={Scatter plots of literacy scores versus attention map similarity for experts (red) and novices (blue) across mini-VLAT, CALVI, and SGL tests. Positive slopes show that higher literacy correlates with expert-like attention patterns.}
    ]{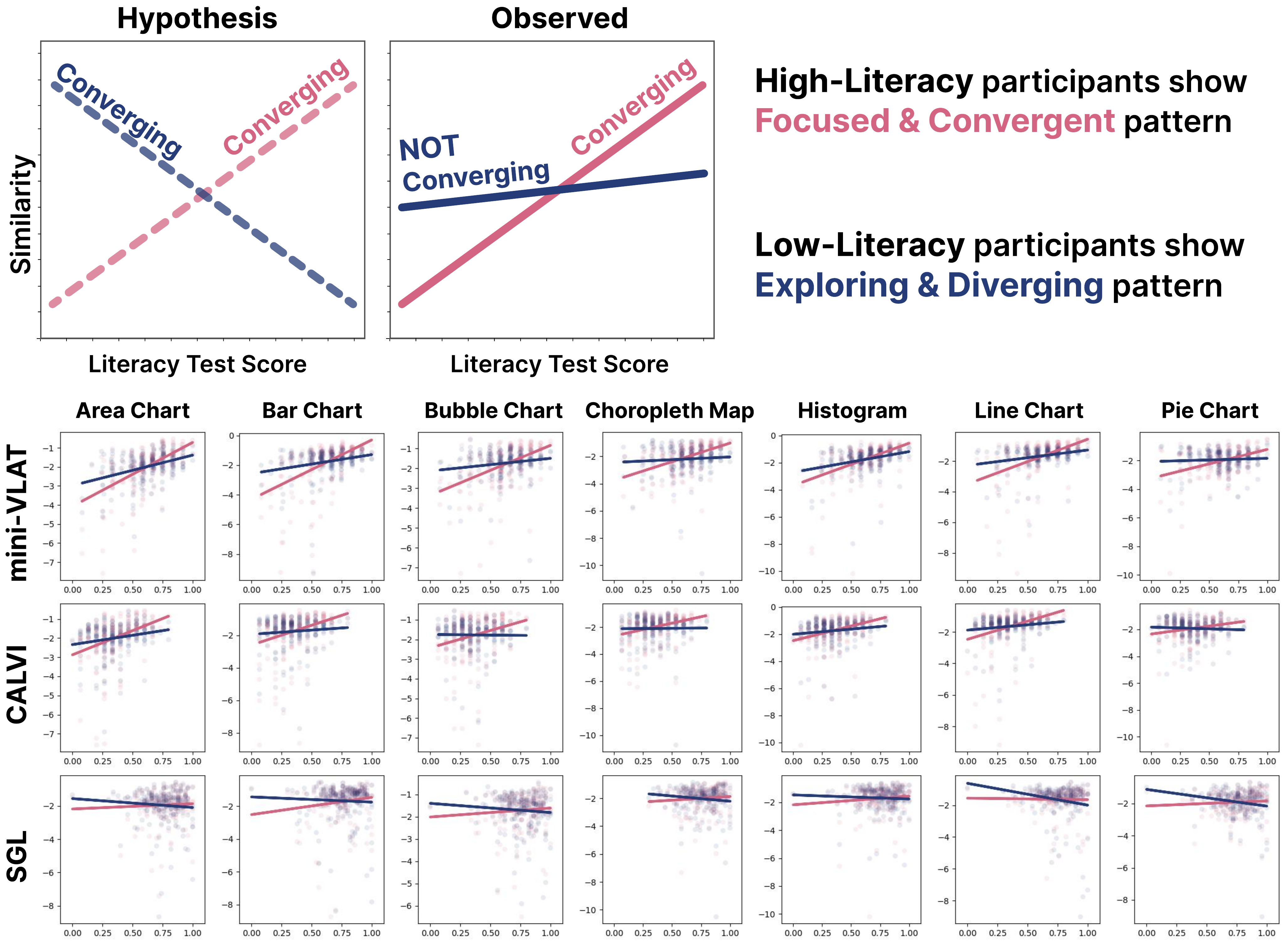}
    \caption{(Top) Description of hypothesis on a similarity in attention map where experts and novices would have common areas in attention. (Bottom) The similarity of individual attention map against two groups calculated with negated KL-D. Only experts' attention converges into a certain pattern. See \autoref{sec:kld} for further analysis.}
    \label{fig:kld}
    \vspace{-5mm}
\end{figure}

\subsubsection{Feature Importance in Linear Regression} \label{sec:linear_regression}
We performed a linear regression to predict literacy scores using these variables.
The results are shown in \autoref{fig:regression}, highlighting the significance and strength of each feature, with all features normalized to a [0, 1] range. Each model for assessment scores partially showed power for explaining the variance in terms of R-squared (mini-VLAT: 0.397, CALVI: 0.341, SGL: 0.273).

On within-map features, click counts had a positive correlation, and task duration had a slightly negative correlation with literacy scores, both highly significant. 
Further analysis of the attention maps showed that experts clicked more frequently in task-relevant, high-focus regions.
This underlines the finding that they focused selectively on fewer areas with greater intensity.

Among the similarity metrics on literacy level groups (experts and novices) of each literacy test,
KL-D and MI had the strongest correlation, showing significance in all literacy types. Other metrics (SRC, SSIM, MSE) had significance on some expertise groups and test scores, but the coefficients were not strong enough compared to the first two. Interestingly, only KL-D and MI negatively correlated with the novice group in all three test scores, showing how defining similarity matters in interpreting viewing patterns.

\subsubsection{Task-Based Analysis of Attention Similarity and Literacy} \label{sec:kld}
Having shown that several features, including KL-D, significantly correlate with literacy scores, we further examine the relationship between literacy and viewing patterns by comparing each participant’s attention map to group-level maps from experts and novices on a per-question basis.
This analysis allows us to test whether participants with similar literacy scores may exhibit convergent viewing patterns. 
We conduct this analysis at the question level because, as shown in~\autoref{sec:result_h1}, some questions or charts may better differentiate between literacy levels, leading to more substantial convergence in viewing patterns among similarly scoring participants.

In \autoref{fig:kld}, we plot the relationship between literacy scores (x) and attention map similarity (y), which describes how strongly the viewing patterns converge to that of the experts (red) and novices (blue). 
We found that \textbf{high-scoring participants tended to show a high similarity of attention pattern with the expert group's map}, showing a positive slope (\includegraphics[width=0.3cm, alt={positively steep pink slope}]{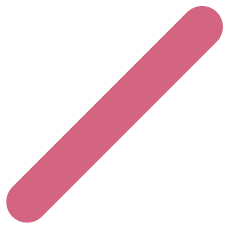}) in all three tests, supporting H3. 
However, \textbf{the attention pattern of low-scored participants did not appear similar to the novice group's}, shown as almost flat slopes (\includegraphics[width=0.3cm, alt={slightly negatively tilted blue slope}]{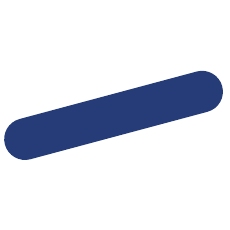}) in mini-VLAT and CALVI tests, providing evidence against H3. 
This further reinforces our findings that participants with high literacy tended to explore the visualization more selectively. In contrast, participants with low literacy tended to explore the image more widely and independently, so the aggregated pattern does not necessarily resemble the individual maps.

\vspace{2mm}
\noindent \textbf{Conclusion:}
We discovered that experts and novices show distinct viewing patterns, quantified by a linear regression test that describes literacy scores with features derived from visual attention.
These results now serve as a foundation for a predictive model that infers how someone views a visualization based on their literacy score.

\begin{figure*}[th]
    \centering
    \includegraphics[
    width=\linewidth,
    alt={Prediction accuracy of Sal2Lit using attention maps from different numbers of charts. Top panel shows line plots for accuracy across literacy levels, while the bottom heatmap ranks chart items by contribution to accuracy.}
    ]{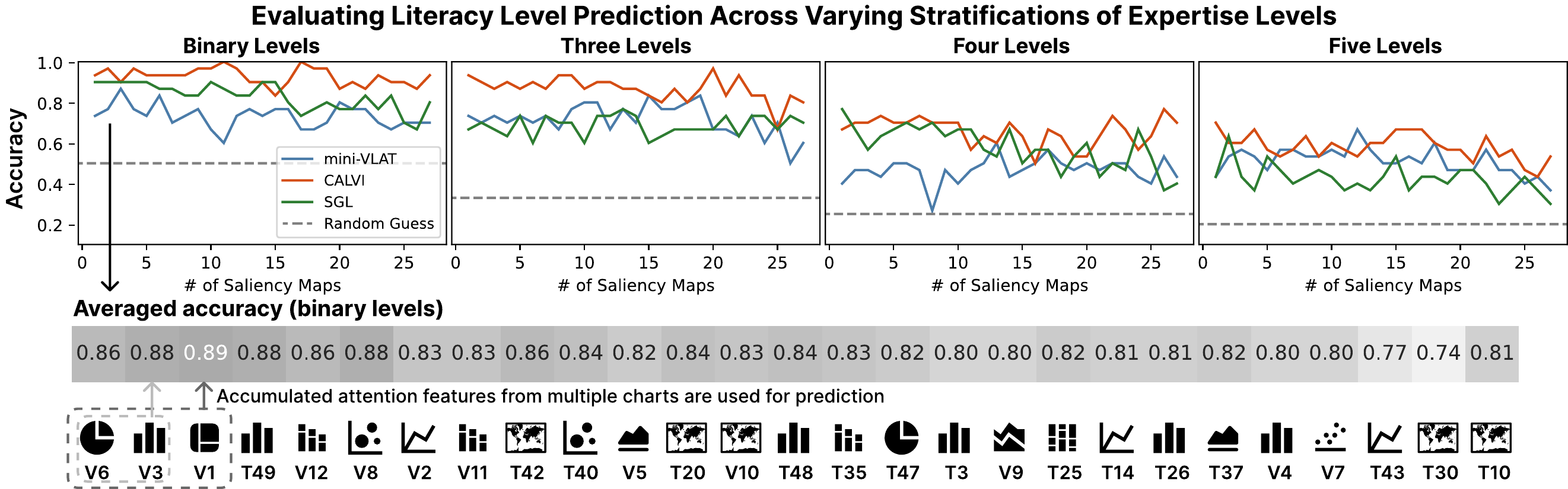}
    \caption{
    Prediction results of participants’ literacy levels using attention map-driven features from one or more charts. The four line charts illustrate prediction accuracy across different numbers of literacy levels, with baseline accuracy indicated by random guessing. The x-axis denotes the number of attention maps used for prediction. The heatmap below details the accumulative features from charts contributing to each performance level. Refer to \autoref{fig:factor} for chart item codes.
    }
    \label{fig:score_prediction}
    \vspace{-5mm}
\end{figure*}
\section{Literacy-Aware Saliency Prediction} \label{sec:saliency_prediction_model}
We trained a saliency prediction model, \textbf{Lit2Sal}, to predict saliency based on task and literacy levels to allow more personalized predictions.

\subsection{Dataset Preparation}
We trained our model with our collected saliency dataset of 235 participants on 27 chart-task pairs with three literacy score types (mini-VLAT, CALVI, SGL).
To ensure a balanced training dataset, we applied oversampling to obtain an equal number of participants from each bin. Specifically, we divided the scores into five quantile-based bins, aligning with the maximum level of expertise differentiation identified in prior work~\cite{expertiseEducation2, dreyfus2004five}.
For evaluation, we formed a test split as 10\% of the full dataset following Wang et al.~\cite{salchartqa}. 
We randomly sampled four participants from each quantile, resulting in 20 (participants) $\times$ 27 (charts) = 540 test data points, accounting for approximately 10\% of the dataset.

\subsection{Model Training}
We created Lit2Sal's model structure by augmenting the original implementation of VisSalFormer~\cite{salchartqa}. We added an information-fusing component in the middle of the model structure to allow the conditional generation of a saliency map based on the literacy score.

\subsubsection{Model Architecture}
We mainly referred to VisSalFormer~\cite{salchartqa} as our baseline structure, the state-of-the-art visualization saliency prediction model. While keeping the Swin-transformer~\cite{liu2021swin} as the image encoder and BERT~\cite{devlin-etal-2019-bert} as the text encoder, we added an additional input of visualization literacy score as a single scalar value.
As shown in the middle panel of \autoref{fig:flow}, the original fusion layer of VisSalFormer~\cite{salchartqa} merged visual (chart-based) and textual information (task-based) via the first two attention modules.
This enables task-focused predictions that can predict viewer attention based on their tasks. 
We appended a second cross-attention module that adds literacy scores to the prediction as a query vector. This allows the literacy score to additionally shape the latent vector.

\subsubsection{Hyperparameters}
We trained Lit2Sal on top of the VisSalFormer's default setting using the Adam optimizer with a learning rate of $5 \times 10^{-5}$ for 300 epochs and a batch size of 128. 
The loss function was calculated with a combination of KL-divergence (KL-D), Normalized Scanpath Saliency (NSS), and Pearson's Correlation Coefficient (PCC), with a ratio of 10:5:2.

\subsection{Evaluation}
We evaluated Lit2Sal’s performance against two state-of-the-art saliency prediction models capable of generating maps that align with human visual attention on images. The first is VisSalFormer~\cite{salchartqa}, our backbone model for task-specific visual saliency prediction. The second is TranSalNet~\cite{lou2022transalnet}, an encoder-decoder Convolutional Neural Network specifically developed to predict saliency that closely follows human visual attention patterns.

\subsubsection{Quantitative Results}
\begin{table}[t]
    \centering
    \resizebox{\linewidth}{!}{
    \begin{tabular}{l S[table-format=1.3] S[table-format=1.3] S[table-format=1.3] S[table-format=1.3] S[table-format=1.3]}
    \toprule
    Model & {PCC ↑} & {NSS ↑} & {AUC ↑} & {SIM ↑} & {KL ↓} \\ 
    \midrule
    TSN (Pre-trained) & 0.156 & 0.714 & 0.677 & 0.154 & 2.689 \\
    TSN (Literacy)    & 0.251 & 1.209 & 0.773 & 0.142 & 2.612 \\
    VSF (Pre-trained) & 0.118 & 0.551 & 0.637 & 0.142 & 2.665 \\ 
    VSF (Literacy)    & 0.347 & 1.500 & 0.833 & 0.188 & 2.327 \\ 
    \midrule
    \textbf{Lit2Sal (mini-VLAT)} & 0.429 & \underline{2.147} & \underline{0.864} & \text{0.255} & \underline{2.100}  \\ 
    \textbf{Lit2Sal (CALVI)} &  \textbf{0.460}   &  \textbf{2.320}  & \textbf{0.874}  &  0.254  & 2.170        \\
    \textbf{Lit2Sal (SGL)} &  \underline{0.430}  &  2.052  &  0.858  &  \textbf{0.257} &  \textbf{2.086}       \\
    \bottomrule
    \end{tabular}
    }
    \caption{
    Model performance comparison on five metrics with pre-trained weights or trained version with our dataset for TranSalNet (TSN) and VisSalFormer (VSF). All metrics for Lit2Sal with all literacy levels have significant differences (p < 0.001 except KL-D: 0.0016, AUC: 0.0022) compared to the VSF model (literacy-trained) based on pairwise t-tests.}
    \label{tab:metrics}
    \vspace{-5mm}
\end{table}

We evaluated the generated saliency maps by measuring the KL-divergence (KL-D), Pearson’s correlation coefficient (PCC), and histogram-based similarity (SIM) to assess their alignment with actual participants’ visual attention maps. 
Because experts and novices produced different attention maps, we assessed their alignment with experts and novices depending on the literacy score input. 
Additionally, we used the Normalized Saliency Scanpath (NSS) and Area Under the Curve (AUC) to quantify how well the generated maps corresponded to human visual attention maps. 
We tested the TranSalNet and VisSalFormer models twice: once with their pre-trained weights and again after fine-tuning them on our literacy dataset. For both models, we averaged the saliency maps of participants with varying literacy scores in the training data, as they were designed to produce literacy-agnostic saliency predictions.

\autoref{tab:metrics} presents the evaluation results, demonstrating that Lit2Sal achieves the highest performance across all metrics for all score types. This result provides strong evidence for literacy-aware saliency predictions. Additionally, we conducted a \textit{t}-test for each metric to compare our model’s performance with the second-best model, the fine-tuned VisSalFormer. Assuming Bonferroni adjustments for multiple-comparisons, 
the analysis revealed the \textbf{statistically significant superiority of the Lit2Sal model across all metrics} (p < 0.001 except KL-D: 0.0016; AUC: 0.0022).
Note that our metric ranges differ from the reported numbers in the original model papers because we treated each participant’s saliency as a single data point without aggregation, which led to more discrete maps and higher KL-D measures.

\subsubsection{Qualitative Results}
The predicted saliency maps from the test set are visualized in \autoref{fig:saliency_prediction}, where expert and novice saliency maps were predicted with top 20\% (first bin) and bottom 20\% (last bin) ranking scores. 
Saliency predictions based on high literacy scores produced more focused maps with fewer noisy areas than those based on low literacy scores. Also, similar to the pattern observed in \autoref{fig:teaser}, we could find that high-literacy people in mini-VLAT and CALVI tend to focus more on the title and axis labels than novices. This could be interpreted as a strategy to solve tricky questions in the CALVI test, where the axis or legends often contain misinformation~\cite{ge2023calvi}.

\vspace{2mm}
\noindent \textbf{Conclusion:}
Our Lit2Sal model empirically maps continuous literacy scores to distinct saliency patterns, serving as a foundational step toward saliency prediction models that account for individual differences.
\section{Visualization Literacy Prediction} 
\label{sec:literacy_level_prediction_model}

Following the successful training of the Lit2Sal model, we conducted another experiment to predict literacy levels using the attention maps, combining VLAT, CALVI, and SGL. We named this model \textbf{Sal2Lit}, where visual attention-driven saliency predicts the literacy level. The diagram of this model is shown in the right panel of \autoref{fig:flow}.

\subsection{Dataset Preparation}
Our dataset size for literacy level prediction is relatively small, consisting of only 235 participants' scores as individual data points. Therefore, we used all the within-map descriptive features (4) and the between-group comparative features (20) as described in~\autoref{sec:result_h3} instead of raw attention maps to support the pattern discovery by our model.

Based on our preliminary experiment, predicting the exact score for all tests was difficult when only visual attention information was used. 
Inspired by use cases where expertise levels were analyzed by groups (see (\autoref{sec:relwork2})), we tested four versions of our Sal2Lit model, each predicting literacy scores as a categorical variable with two to five levels.
The levels were determined based on score distribution quantiles. 
This allowed us to balance the number of data points in each quantile with oversampling, leading to more robust predictions. Also, following the evaluation setup of Sal2lit, we sampled the same 20 participants (four from each quantile) for the test set.

\subsection{Model Training}
Our Sal2Lit model was trained to predict three classes of literacy scores: mini-VLAT (basic comprehension), CALVI (critical thinking), and SGL (self-assessment). 
We also used a greedy algorithm~\cite{cormen2009introduction} to identify the literacy test chart(s) that generate the most predictive visual attention maps, optimizing accuracy and minimizing data usage.

\subsubsection{Model Architecture \& Hyperparameters}
Based on our preliminary experiments, traditional machine learning models such as Random Forest and Gradient Boosting proved insufficient for capturing the complexity of the problem. To address this, we employed a deep learning approach using a feedforward neural network with stacked linear layers, enabling the model to learn hierarchical feature representations.

Sal2Lit's network structure consists of five layers, with the hidden dimension starting at 512 and progressively halving until it reaches 32 before mapping to the final output layer corresponding to the number of classes (literacy levels). 
Our user study result showed that literacy test scores are correlated. Therefore, we aimed to predict all three literacy test levels simultaneously with a single model. To balance the prediction of the three levels, we averaged three individual cross-entropy loss functions with equal weights.
We used the Adam optimizer with a learning rate of $1 \times 10^{-4}$ for optimization. A batch size of 256 was chosen to maximize the GPU memory usage.
The model was trained for a maximum of 150 epochs, with early stopping implemented to mitigate overfitting.

\subsubsection{Greedy Subset Selection for Influential Charts} \label{sec:greedy}

We aim for our visual attention-based model to supplement literacy tests by efficiently predicting literacy using only a few charts, minimizing time and effort.
Exhaustively evaluating all possible chart item combinations is computationally expensive, so we applied a greedy selection algorithm to approximate the optimal subset. 
In this approach, we first identified the single chart item with the highest prediction accuracy when used alone. 
This was the bar chart from the mini-VLAT (V6), as shown in \autoref{fig:score_prediction}.  
We then iteratively added one chart item at a time, each time selecting the item that produced the most significant improvement in prediction accuracy when combined with the previously selected items. Accuracy was measured as the average across the three literacy tests. This process continued until every chart item was added. Through this strategy, we efficiently identified a high-performing set of charts for literacy prediction without evaluating all possible combinations.

We also leave how to average three literacy scores' accuracy as a controllable parameter. For example, in some cases, the literacy level of comprehension (VLAT) would be preferred over critical thinking (CALVI), where all charts are proven to have no misinformation. Based on these scenarios, we can apply a weighted average to prioritize VLAT over CALVI and SGL.

\subsection{Evaluation}
We evaluated the trained Sal2Lit model based on the accuracy of its prediction results. The summarized results are plotted in \autoref{fig:score_prediction}.

\subsubsection{Overall Accuracy}
Matching the intuition, the accuracy for binary classification of literacy level was generally higher than on more diverse levels in all three tests. Our model could predict the binary literacy level with \textbf{an average accuracy of 86\% in all three literacy levels using only one visual attention map} (mini-VLAT: 73\%, CALVI: 93\%, SGL: 90\%). 
Moreover, the accuracy exceeded 87\% across all three tests using only three visualizations' attention maps: critical thinking (CALVI) reached 97\%, self-assessment (SGL) 90\%, and basic comprehension (mini-VLAT) 87\%. However, contrary to the expectation that incorporating information from additional charts would enhance performance, 
accuracy did not improve significantly with added charts and visual attention maps.
This indicates a limit to the meaningful information that can be extracted from visual attention maps, beyond which the model begins to learn noise rather than meaningful patterns.

Comparing the results among the literacy types, we observe that CALVI’s prediction accuracy is almost always higher than mini-VLAT and SGL. This suggests that the \textbf{difference in critical thinking ability is shown most vividly} as the difference in the attention pattern. In comparison, the mini-VLAT's accuracy was significantly lower than the other two tests, suggesting that the \textbf{general comprehension level is more difficult to capture} only based on the visual attention patterns.

The optimal charts and tasks selected by the greedy algorithm are shown in the bottom of \autoref{fig:score_prediction}, which illustrates the ranking of the charts that contributed most to distinguishing experts from novices in literacy level prediction based on visual attention maps. 
The top three items (V6, V3, V1) with the highest average accuracy were from the mini-VLAT test. 
This suggests that the critical thinking ability measured by CALVI, can be captured even through VLAT questions without relying on trick questions. 
We discuss this deeper in~\autoref{sec:discussion}.

\subsubsection{Analyzing Feature Importance} \label{sec:sal2lit_importance}

To support our trained Sal2Lit model's robustness of its prediction performance, we analyzed which feature (from \autoref{sec:result_h3}) contributed the most to the model's prediction. We applied the Integrated Gradient (IG)~\cite{mudrakarta2018did} technique to measure how each feature strongly affects the gradient flow. The impact is calculated by aggregating the gradient value in our model $F$, starting from the baseline feature $V_0=0$ to our test data's feature $V$. This is formulated as: 
\vspace{-1mm}
\begin{equation}
\mathbf{IG}(\mathbf{V}) = \mathbf{V} \int_{0}^{1} \nabla_{\mathbf{V}} F(\alpha \mathbf{V}) \, d\alpha
\label{eq:integrated_gradient}
\vspace{-1mm}
\end{equation}
The result of applying IG to the binary-level prediction model is plotted in \autoref{fig:regression} as a blue color scale of the predictive model. 
Among \textbf{the within-map descriptive features} (see \autoref{sec:relwork2}), click counts had the highest regression coefficient (computed from our user study results), but Shannon entropy had the highest feature importance (computed from our Sal2Lit model) despite a low regression coefficient. 
This discrepancy between empirical user-driven results and model-based interpretation suggests a non-linear relationship of entropy value predicting literacy.
Among \textbf{between-group comparative features}, KL-D showed both high feature importance and regression coefficients. 
Note that the calculated feature importance reflects the behavior of the trained model and does not necessarily indicate generalizable knowledge.

\vspace{2mm}
\noindent \textbf{Conclusion:}
Our Sal2Lit model accurately predicts visualization literacy using features from as few as three visual attention maps, validating our hypotheses that viewing patterns can effectively distinguish between varying levels of literacy.
\section{Discussion} \label{sec:discussion}
Several findings in our study offer new insights into how visual attention reflects literacy and critical thinking.

\vspace{1mm}
\noindent\textbf{Implications for adaptive visualization and scaffolding.} 
Our study reveals diverging behavior among literacy levels: experts show focused, `convergent' attention on contextual elements like titles, while novices display dispersed, divergent' visual exploration. 
We reflect on our findings to consider their implications for adaptive visualization design.

While our current findings do not allow for any causal conclusions about whether novices become disoriented within the visualization or whether the visualization itself fails to guide them, we propose two plausible explanations. 
On one hand, novices may be more susceptible to cognitive overload when faced with unfamiliar visual structures, leaving them uncertain about where to begin extracting information. 
On the other hand, the visualization might not be designed to help certain viewers locate key patterns, increasing the likelihood of incorrect responses on comprehension or critical thinking questions. 
In both scenarios, an adaptive system that detects disoriented viewing patterns could intervene with visual cues, like highlighting axes or annotations, to support user understanding.
However, we acknowledge the risk that such designs may reduce user agency.
As every visualization is a rhetorical device with multiple possible interpretations~\cite{bearfield2024same}, adaptive systems must be designed with careful attention to the story the visualization intends to convey.

\vspace{1mm}
\noindent\textbf{Redefining critical thinking in visualization.} 
In our experiment, viewing patterns on mini-VLAT charts, which contained no misinformation, were strongly predictive of CALVI test scores. 
This finding suggests the presence of expert-like viewing behaviors associated with critical thinking, particularly those involving scrutiny of a chart’s structure. 
These results point to the potential for a novel class of literacy assessments: rather than relying solely on standardized tests, researchers could evaluate critical thinking by analyzing viewing patterns on visualizations, offering a means to detect skepticism (and potentially even perceived credibility) through visual behavior.

\section{Limitations \& Future Work} \label{sef:limitation}

We trained models to differentiate visualization novices from experts using visual attention maps from visual analytic tasks.
However, our data collection method of BubbleView~\cite{bubbleView} is only a proxy of visual attention. 
BubbleView requires slow, deliberate clicking at a rate far slower than natural eye movements.
This may disrupt natural visual exploration and increase cognitive load, reducing response accuracy~\cite{shin2017effects}.
Future work can test our models with eye-tracking data~\cite{webgaze} and fine-tune the outcome to reveal a deeper relationship between attention patterns and literacy. 

Further, visual literacy can be multi-faceted. 
Our current experiment tested only three tests: mini-VLAT, CALVI, and SGL.
Future work can improve our seminal work by training the models on more extensive chart datasets with more participants that cover a broader spectrum of tasks and literacy levels. Such expanded evaluations would increase the performance and generalizability of our model across varied contexts

Finally, we found that similar visual attention patterns do not always indicate the same answers on a literacy test (see \autoref{fig:correctness} Pie Chart).
Participants can focus on the same regions, producing both correct and incorrect responses.
This suggests that attention does not guarantee comprehension, as misinterpreted prompts or time pressure can lead to errors~\cite{burton2005multiple, onwuegbuzie1995effect}. 
This highlights the limitations of relying solely on visual attention as a proxy for cognitive processing and visual literacy. 
Future work should explore additional diagnostic cues that more accurately reflect literacy levels. 
Therefore, we claim that our \textbf{Sal2Lit model complements, rather than replaces prior literacy assessments.}
\section{Conclusion}
In this work, we conducted a user study investigating the relationship between visualization literacy and human visual attention patterns. 
Our results revealed distinct human visual attention patterns based on different literacy levels: experts selectively focused on fewer regions but with greater intensity, while novices explored more broadly.
Building upon these findings, we introduced two computational models: \textbf{Lit2Sal}, a literacy-aware saliency model predicting where people will attend to in a visualization depending on their literacy levels, and \textbf{Sal2Lit}, which predicts one's visualization literacy level from their visual attention maps. 
Together, these models open opportunities for real-time adaptive visualizations that account for individual differences by customizing visuals to match viewers’ literacy. 
Also, our approach has promising educational applications, such as attention-based feedback mechanisms that guide novices to attend to task-relevant regions. 
Finally, our models show potential for integration into design recommendation tools that dynamically evaluate and suggest improvements during the visualization authoring process, enhancing clarity and comprehension for audiences with diverse literacy levels.

\vspace{-2mm}
\acknowledgments{%
    This work was supported in part by the National Science Foundation awards IIS-2237585, IIS-2311575, and III-2453462. 
  Y. Wang was funded by the Deutsche Forschungsgemeinschaft (DFG, German Research Foundation) -- Project-ID 251654672 -- TRR 161. Supplemental materials are accessable in https://osf.io/2crb9
}
\vspace{-2mm}

\bibliographystyle{abbrv-doi-hyperref}

\bibliography{template}

\appendix 

\end{document}